\documentclass[twocolumn,letter]{jpsj2}
\title{A New Scenario on the Metal-Insulator Transition in VO$_2$}

\author{
Arata {\sc Tanaka}
}

\inst{
Department of Quantum Matter, ADSM\\
 Hiroshima University, Higashi-Hiroshima 739-8530, Japan
}

\recdate{ }

\abst{
The metal-insulator transition in VO$_2$ 
was investigated using the three-band Hubbard model,
in which the degeneracy of the 3$d$ orbitals, the on-site Coulomb and exchange
interactions, and the effects of lattice distortion were considered.
A new scenario on the phase transition is proposed, where the
increase in energy level separation among the $t_{2g}$ orbitals caused
by the lattice distortion triggers an abrupt change in the electronic configuration 
in doubly occupied sites from an $S=1$ Hund's coupling state to a spin $S=0$ 
state with much larger energy, and this strongly suppresses the charge fluctuation.
Although the material is expected to be a Mott-Hubbard insulator in the insulating phase,
the metal-to-insulator transition is not caused by an increase in relative strength of the Coulomb interaction
against  the electron hopping as in the usual Mott transition,
but by the level splitting among the $t_{2g}$ orbitals against the on-site exchange interaction.
The metal-insulator transition in Ti$_2$O$_3$ can also be explained by the same scenario.
Such a large change in the 3$d$ orbital occupation at the phase transition can be detected by linear dichroic V 2$p$ 
x-ray absorption measurements. 
}
\kword{
VO$_2$, metal-insulator transition, Mott-Hubbard insulator, exchange interaction, lattice distortion, Ti$_2$O$_3$, V 2$p$ XAS
}

\begin{document}
\maketitle
\begin{figure}
\begin{center}
\includegraphics[width=5.0cm]{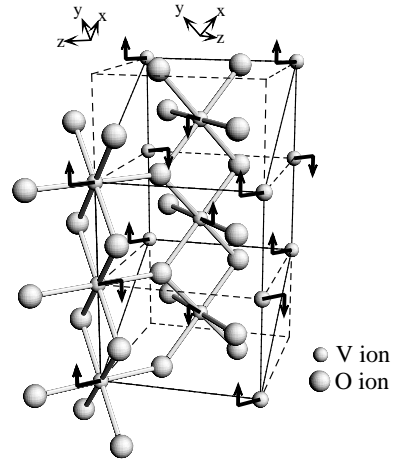}
\end{center}
\caption{Crystal structure of VO$_2$. The large and small circles denote oxygen and V ions, respectively.
The unit cell in the monoclinic ($M_1$) phase is indicated by solid lines and those in the rutile ($R$) phase 
by dashed lines. The arrows indicate exaggerated displacement of the V ions in the $M_1$ phase.}\label{VO2cry}
\end{figure}

Vanadium dioxide undergoes a metal-insulator transition (MIT) at $T=340$~K\cite{Imada}.
Simultaneously, a lattice distortion from the high-temperature rutile ($R$) structure
takes place, and  in the low-temperature monoclinic ($M_1$) phase, V ions pair and the unit cell doubles. 
Each V ion (nominally V$^{4+}$ with 3$d^1$ configuration)
in the VO$_2$ crystal is surrounded by an octahedron consisting of six nearest-neighbor 
oxygen ions (see Fig.~\ref{VO2cry}) and an octahedral crystal field splits the 3$d$ levels into twofold $e_g$ and threefold $t_{2g}$ levels.
The energies of the low-lying $t_{2g}$ orbitals are further differentiated by a small $D_{2h}$ crystal field
and the orbitals can be described as $|\sigma\rangle=|xy\rangle$, 
$|\pi 1\rangle=\frac{1}{\sqrt{2}}\{|yz\rangle+|zx\rangle\}$
and $|\pi 2\rangle=\frac{1}{\sqrt{2}}\{|yz\rangle-|zx\rangle\}$ 
using $x$-, $y$- and $z$-coordinates locally defined for each V ion, as shown in Fig.~\ref{VO2cry}.
In the rutile phase, V ions form a linear chain along the $c$-axis, and oxygen octahedrons surrounding the V ions share their edges in each chain. Two kinds of chains having the shared edges along $[110]$ and $[1\bar{1}0]$ directions are
present.
While the $\sigma$ orbitals have lobes along the chain and are strongly hybridized with their neighbors on the chain,
the coupling among the $\pi$ orbitals on the chain is weak. 
In the $M_1$ phase,  V ions on each chain form pairs and the V-V bonds of the pairs tilt from the $c$-axis.

A number of theories based on Peierls\cite{Goodenough,Wentzconvitch,Eyert}, Spin-Peierls\cite{Paquet} and Mott-Hubbard\cite{Zylbersztejn,Korotin} mechanisms  have been proposed for the MIT.
Since V ions form pairs in the $M_1$ phase, the Peierls mechanism appears to be
a good candidate as an explanation for the MIT. However, experimental results obtained from studies on the Cr-doped alloy V$_{1-x}$Cr$_x$O$_2$\cite{Pouget} and on VO$_2$ under uniaxial stress\cite{Pouget2}
are inconsistent with such an interpretation.
In addition to the $M_1$ and $R$ phases, insulating monoclinic ($M_2$) and triclinic ($T$) phases at the temperature between the $M_1$ and $R$ phases are found in Cr-doped samples and VO$_2$ under pressure.
In the $M_2$ phase, V ions on one-half of the chains pair without the tilting of their V-V bonds,
whereas those on the other half form zigzag chains without pairing.  These zigzag chains with equally spaced V ions
behave magnetically as $S=1/2$ Heisenberg chains with $J\sim 300$~K and are considered to be Mott-Hubbard insulators.
Furthermore, a continuous $M_2$-to-$M_1$ transition through 
the intermediate $T$ phase is observed as temperature decreases.
Since these two insulating phases appear by doping a small amount of Cr or uniaxial pressure,
all the $M_1$, $M_2$ and $T$  phases are classified as Mott-Hubbard and not band insulators\cite{Rice}.

The purpose of this letter is to discuss the mechanism of the MIT in VO$_2$  
 using the three-band Hubbard model on the basis of the many-body theory,
particularly in relation to the on-site exchange interaction and the level splitting among the $t_{2g}$ orbitals caused by the lattice distortion. 

As a model for the V ion chain along the rutile $c$-axis in VO$_2$, 
a linear chain cluster consisting of six V ions with the periodic condition is assumed
and the three $t_{2g}$ orbitals $\sigma$, $\pi 1$ and $\pi 2$ are considered in each V ion.
The Hamiltonian assumed in the model is written as 
\begin{equation}
 H_\textrm{elec}=H_\textrm{cry}+H_\textrm{hyb}+H_\textrm{int}.\label{Ham}
\end{equation}
In the band structure calculations, the energy band arising from the $\pi$ orbitals 
is shifted to higher energy due to the pairing and zigzag distortions of
the V ion chains in the insulating phases\cite{Goodenough,Eyert}. 
To consider such effects of the lattice distortions in the $M_1$ and $M_2$ phases, 
a low-symmetry crystal field $H_\textrm{cry}$ which splits the
$\sigma$ and $\pi$ orbital levels is introduced and is described as
\begin{equation}
H_\textrm{cry}=\frac{D}{3}\sum_i \left(n_{i,\pi 1}+n_{i,\pi 2}-2n_{i,\sigma}\right),
\end{equation}
where $D$ is the energy difference between the $\sigma$ and $\pi$ orbitals and 
$n_{i,m}$ is the number operator for an orbital $m$($=\sigma$, $\pi 1$ or $\pi 2$) at the $i$-th V ion site.

$H_{\rm hyb}$ represents electron hopping between the nearest neighbor V ion sites and is written as
\begin{equation}
    H_\textrm{hyb}=\sum_{i,m,s}V_m\left(a^\dagger _{i,m,s}a^{}_{i+1,m,s}+a^\dagger_{i+1,m,s}a^{}_{i,m,s}\right),
\end{equation}
where $a^\dagger_{i,m,s}$ denotes the creation operator for an electron 
on an orbital $m$ with a spin quantum number $s$ at the $i$-th V ion site and $V_m$ is the hopping integral.
For simplicity, the variations in the values of the hopping integrals arising from the pairing and zigzag distortions
of the chains in the insulating phases are not considered.
However, even if these effects are included, there is no qualitative change in the results obtained.
Thus, $D$ is the only essential parameter of the distortion for the mechanism of the MIT,
and the Peierls and Mott-Hubbard mechanisms triggered by the change in the strength of the hybridization cannot be the main cause of the MIT within the present model.

The last term $H_{\rm int}$ stands for the on-site Coulomb and exchange interactions: 
\begin{equation}
\begin{split}
   H_\textrm{int}&=U\sum_{i,m}n_{i,m\uparrow}n_{i,m\downarrow} + \frac{U'}{2}\sum_{i,m\neq m'} n_{i,m}n_{i,m'}\\
   &-\frac{J}{2}\sum_{i,m \neq m' \atop s,s'}a^\dagger_{i,m,s}a^{}_{i,m,s'}a^\dagger_{i,m',s'}a^{}_{i,m',s} \\
   &-\frac{J}{2}\sum_{i,m \neq m' \atop s,s'}a^\dagger_{i,m,s}a^{}_{i,m',s'}a^\dagger_{i,m,s'}a^{}_{i,m',s},
\end{split}
\end{equation}
where $U$ and $U'$ are the Coulomb repulsion energies between electrons on the same and different orbitals, respectively, and $J$ is the exchange interaction energy; there exists a relation $U=U'+2J$ among them.
The values for $U$, $U'$ and $J$ are determined from high-energy spectroscopic data.
Note that although only the V 3$d$ orbitals are 
included in the present model, the hybridization between the 3$d$ and neighboring oxygen 2$p$ orbitals 
is  strong in the actual system. 
To consider the effects of the 3$d$--2$p$ mixing, the parameter values were adjusted to reproduce a low-energy level diagram obtained from a cluster model, in which both the V 3$d$ and O 2$p$ orbitals are included\cite{Tanaka}.
For the hopping integrals between the $t_{2g}$ orbitals,
not only the direct $d$-$d$ hopping but also those via neighboring oxygen 2$p$ orbitals
is also considered within the second-order perturbation theory: 
$V_\sigma=(3dd\sigma +dd\delta)/(4N_\pi)$, $V_{\pi 1}=(dd\pi-pd\pi^2/\Delta)/N_\pi$ and
$V_{\pi 2}=(dd\delta+pd\pi^2/\Delta)/N_\pi$.
Here, $dd\sigma$, $dd\pi$ and $dd\delta$ are the Slater-Koster parameters 
for the 3$d$ orbitals between the nearest neighbor V ions, $pd\pi$ is that between
the 3$d$ and 2$p$ orbitals, $\Delta$ is the charge transfer energy between the 3$d$ and 2$p$ orbitals and $N_\pi=1+4(pd\pi/\Delta)^2$.
For the values of $\Delta$ and $pd\pi$, those obtained from analysis of the high-energy spectroscopy experiments are
used, and for $dd\sigma$, $dd\pi$ and $dd\delta$, Harrison's values are adopted\cite{Harrison}.
The parameters adopted in the model are $U'=1.33$~eV, $J=0.65$~eV, $V_\sigma=-0.25$~eV, 
$V_{\pi 1}=-0.04$~eV and $V_{\pi 2}=0.22$~eV. 
The ground-state and low-lying excited-states eigenvalues and eigenvectors were calculated 
by numerical diagonalization of $H_{\rm elec}$ using the Lanczos method.

A possible scenario on the MIT is as follows. When the value of $D$
is small, the electronic configuration $\sigma \pi$ (one of the two electrons occupies the $\sigma$ orbital and 
the other the $\pi 2$ orbital) with total spin $S=1$ for the doubly occupied site is favorable because of Hund's first rule, and this causes nearest-neighbor ferromagnetical antiferro-orbital correlation as $\sigma\!\!\uparrow$--$\pi\!\!\uparrow$.
On the other hand, for a large $D$ compared to $J$, the $S=0$ $\sigma \sigma$ configuration is favorable 
for the doubly occupied site and this induces nearest-neighbor antiferromagnetical ferro-orbital
correlation as $\sigma\!\!\uparrow$--$\sigma\!\!\downarrow$.
Because on-site electron repulsion energy for the $S=1$ $\sigma \pi$ configuration
is $U'-J=0.7$~eV whereas that for the $S=0$ $\sigma \sigma$ configuration
is $U= 2.6$~eV, charge fluctuations $d^1;d^1\rightarrow d^0;d^2$ should be strongly suppressed
for large $D$. 

\begin{figure}
\begin{center}
\includegraphics[width=7.0cm]{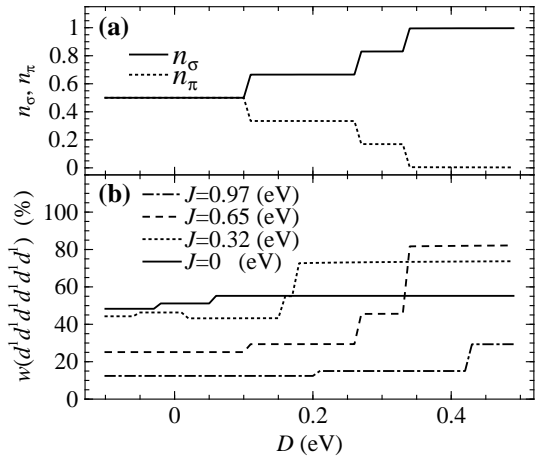}
\end{center}
\caption{$D$ dependence of the $\pi 2$ and $\sigma$ orbital occupation per V ion (a) and weight of  
the $d^1d^1d^1d^1d^1d^1$ configuration (b) in the ground state
of the V ion chain model.  For the weight, results obtained with various values of $J$ are depicted.}\label{v6pDdep}
\end{figure}
Figure \ref{v6pDdep} shows $D$ dependence of the $\pi 2$ and $\sigma$ orbital occupation
 per V ion (a) and the weight of  the $d^1d^1d^1d^1d^1d^1$ configuration (b) in the ground state
of the V ion chain cluster model. To determine the effects of $J$ on the charge fluctuation, in Fig.~\ref{v6pDdep}(b),
the weights obtained with various values of $J=0$, $0.32$, $0.65$ and $0.97$~eV are shown.
As anticipated in the above discussion, for $D< 0.1$~eV,
the ratio of the orbital occupation is $n_\sigma:n_\pi=1:1$, which implies
the presence of the $S=1$ $\sigma\pi$ doubly occupied sites and
$\sigma\!\!\uparrow$--$\pi\!\!\uparrow$ correlation in the
ground state, and for $D> 0.34$~eV, the ratio is $n_\sigma:n_\pi=1:0$, which indicates 
the presence of the $S=0$ $\sigma\sigma$ doubly occupied sites and $\sigma\!\!\uparrow$--$\sigma\!\!\downarrow$ correlation.
Since we find other states with intermediate orbital occupations
$n_\sigma:n_\pi=2:1$ and $n_\sigma:n_\pi=5:1$ in $0.10 < D <0.34$, the variation from the state with the 
$S=1$ $\sigma\pi$ to $S=0$ $\sigma\sigma$  doubly occupied sites with
increasing $D$ is probably continuous in the infinite-sites limit.
In Fig.~\ref{v6pDdep}(b), we can find an approximately 60\% increase in the weight of the $d^1 d^1 d^1 d^1d^1d^1$
configuration from the ground state around $D=0$~eV to that around $D=0.4$~eV 
with $J=0.65$~eV. This indicates a large change in the charge fluctuation between the two  states.
As discussed earlier, this change in the charge fluctuation originates from the difference in the  
electron repulsion energy of the doubly occupied sites: $U'-J$ for the $S=1$ $\sigma \pi$ configuration
and $U(=U'+2J)$ for the $S=0$ $\sigma \sigma$ configuration. Indeed, both the difference in the weights
and the values of $D$ where the change in the configurations take place
 increase with  increasing $J$, showing that the on-site exchange interaction is essential 
for the MIT. 

The two kinds of electronic configurations discussed above are analogous to the so-called high-spin and low-spin states
in a transition metal ion embedded in an octahedral crystal field scaled by the energy difference between 
the $e_g$ and $t_{2g}$ orbitals $10Dq$.
If the value of $10Dq$ is small with respect to the intra-atomic exchange interaction energy,
 the 3$d$ electronic state still has the same total spin
as expected from Hund's first rule and is called a high-spin state, whereas for 
the crystal field superior to the exchange interaction, an electronic state with all electrons 
in low-lying $t_{2g}$ orbitals having smaller total spin is favorable and is called a low-spin state.
From this viewpoint, the phase transition can be interpreted as a ``high-spin'' $S=1$ $\sigma \pi$ state
to ``low-spin'' $S=0$ $\sigma \sigma$ state transition in doubly occupied
sites  caused by a change in the crystal field $D$.

\begin{figure}
\begin{center}
\includegraphics[width=8.0cm]{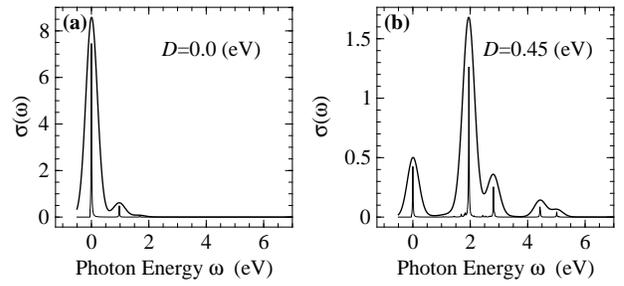}
\end{center}
\caption{Optical conductivity spectra calculated with $D=0$ eV (a) and $D=0.45$ eV (b).}\label{v6pOcd}
\end{figure}
Optical  conductivity spectra were calculated to determine the difference in electronic transport properties
between the ground states with the low-spin and high-spin doubly occupied sites
corresponding to the insulating and metallic phases in real material, respectively.
The method described in ref.~\citen{Dagotto} is employed for the calculation with the six V ion cluster model.
Figure \ref{v6pOcd} shows the optical conductivity spectra as a function of the photon energy $\omega$
calculated with $D=0$~eV (a) and $D=0.45$~eV (b). The spectra were convoluted with the Gaussian distribution curve with two different broadening widths 0.02 eV and $0.4$ eV.
As seen in Fig.~\ref{v6pOcd}(a), a large portion of the intensity $\sim$95\% is concentrated at the Drude peak,
which is positioned at $\omega =0$, and this clearly shows the metallic feature of the  ground state with the $S=1$ $\sigma \pi$ configuration in the doubly occupied sites. 
On the other hand, the spectrum for the ground state with the $S=0$ $\sigma \sigma$ configuration 
 in Fig.~\ref{v6pOcd}(b) has a massive peak at $\sim$2~eV and the intensity of the Drude peak is reduced to only $\sim 5$\% from that in Fig.~\ref{v6pOcd}(a). Although the weak Drude peak still remains, probably due to the finite size effect, such a large reduction in its intensity indicates the localization of electrons in the ground state with the low-spin doubly occupied sites.

\begin{figure}
\begin{center}
\includegraphics[width=7.0cm]{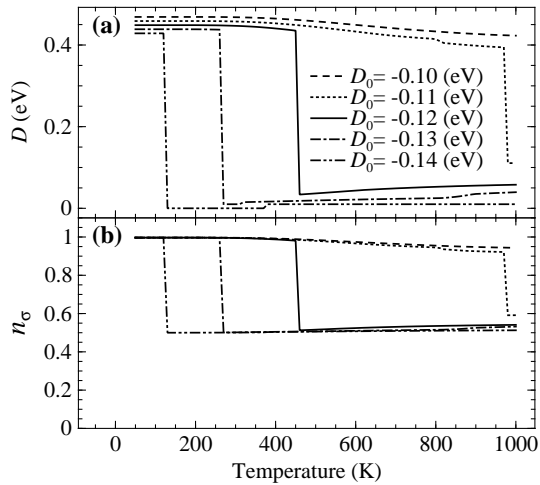}
\end{center}
\caption{Temperature dependence of $D$ which minimizes the free energy $F(T;D)$ (a)
and the expectation value of $n_\sigma$ (b) obtained with $K=1.17$ (eV)$^{-1}$ and various values of $D_0$.
}\label{v6pTdep}
\end{figure}
We have discussed the variation in the electronic structure caused by the monoclinic lattice distortion scaled by $D$.
In the real material, the first-order MIT takes place as temperature decreases accompanied with the lattice distortion.
To demonstrate how the change in the electronic structure as a function of $D$
can be related to the temperature-induced MIT in  VO$_2$, let us consider a simple model for the lattice elastic energy, where the energy per
V ion site is a quadric function of  $D$: $\frac{K}{2}(D-D_0)^2$.
The free energy $F(T;D)$ of the system then can be written as 
\[
F(T;D)=-k_\textrm{B}T\log Z_\textrm{elec}+N_\textrm{site}\frac{K}{2}(D-D_0)^2,
\]
where $Z_\textrm{elec}$ is the partition function of $H_\textrm{elec}$ defined in eq.~(\ref{Ham}) and 
the second term is the lattice elastic energy and
$N_\textrm{site}=6$ is the number of V ion sites in the model.
Figure \ref{v6pTdep} shows values of $D$ which minimize the free energy $F(T;D)$ for each $T$
calculated with $K=1.17$~(eV)$^{-1}$ and $D_0=-0.10$, $-0.11$, $-0.12$, $-0.13$  and $-0.14$~(eV).
In Fig.~\ref{v6pTdep}(b), the expectation values of the $\sigma$ orbital occupation are also depicted.
In Fig.~\ref{v6pTdep}(a), a clear discontinuous change in $D$ as a function of $T$
can be seen. With this jump in $D$, the electronic state
also changes from the localized state with the $S=0$ $\sigma\sigma$ sites
to the delocalized one with the $S=1$ $\sigma\pi$ sites, which is evident from Fig. ~\ref{v6pTdep}(b).
This temperature-induced abrupt transition, both in the lattice distortion 
and the electronic state, which is consistent with the experimental MIT, can be understood as follows.
For the state with $S=1$ $\sigma\pi$ doubly occupied sites, the excitation originated from 
 its spin, and orbital degrees of freedom  positioned densely within the low-energy region $<0.1$~eV
lower the free energy at high temperatures. On the other hand, such an energy lowering by the entropy term 
is minor for the state with the $S=0$ $\sigma\sigma$ sites, since the state has only spin excitations.

\begin{figure}
\begin{center}
\includegraphics[width=7.0cm]{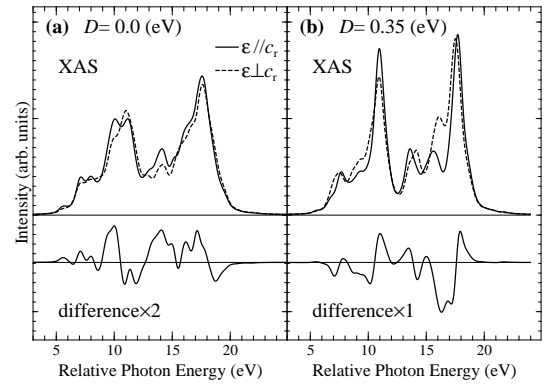}
\end{center}
\caption{V 2$p$ XAS spectra for two different directions of the polarization vectors $\mib{\varepsilon}$, which are parallel (solid line) and perpendicular (dashed line) to the rutile $c$-axis, calculated with $D=0$ eV (a) and $D=0.35$ eV (b). The difference spectra are shown at the bottom in each panel.}\label{V4ld}
\end{figure}
If there is a large change in the electronic configuration between the metallic and insulating phases, 
as is predicted in the present theory, such a difference can be detected by measuring the linear dichroism in the
V $2p$ x-ray absorption spectrum (XAS), since the spectrum is very sensitive to the 3$d$ orbital occupation.
To show this, the spectra were calculated using a cluster model with four V ions, where, in addition to 
the $t_{2g}$ orbitals\cite{Tanaka,Tanaka2}, 
the $e_g$ orbitals are considered for each V ion site.  While the ground state properties are almost unchanged, 
inclusion of the $e_g$ orbitals is important for the final state of the spectra.
In the final state, the 2$p$ core hole of photoexcited site, and the Coulomb and exchange interactions between 
the 3$d$ electrons and the 2$p$ core-hole are further taken into account.
Figure \ref{V4ld} shows the theoretical spectra obtained with $D=0.0$~eV (a) and $D=0.35$~eV (b) corresponding to 
those for the metallic and insulating phases, respectively.
The spectra are depicted for two different directions of the polarization vector $\mib{\varepsilon}$ of the light:
parallel and perpendicular to the rutile $c$-axis. The difference spectra are also shown at the bottom in each panel.
Because of the selection rule of the $2p \to 3d$ dipole transition, the transition probability to the $\sigma$ orbital is larger than that to the $\pi$ orbitals for $\mib{\varepsilon} /\!/ c_r$, whereas
that to the $\pi$ orbitals is  larger  for the $\mib{\varepsilon} \perp c_r$. 
Indeed, the integrated intensity of the difference spectra for the ``metallic'' ground state
with the orbital occupation $n_\sigma:n_\pi$=$1:1$ and the ``insulating'' ground state
with $n_\sigma:n_\pi$=$1:0$ have mutually opposite signs,
in addition to the large differences in the spectral shape itself.

An MIT in Ti$_2$O$_3$ can also be
explained by the same mechanism proposed here.
Ti$_2$O$_3$ exhibits a gradual MIT around 400--600~K and has the corundum structure,
where Ti ions form pairs along the $c_h$-axis\cite{Imada}.
Because of a small trigonal field, the $t_{2g}$ levels split into
low-lying $a_{1g}$ and twofold ${e_g}^\pi$ levels.
At high temperature, the electronic configuration of the
doubly occupied sites is expected to be $S=1$ $a_{1g}{e_g}^\pi$.
Although the symmetry of the crystal is not changed, the Ti-Ti bonds of Ti ion pairs become shorter and the energy separation of the ${e_g}^\pi$ and $a_{1g}$ orbitals
increases as temperature decreases, and this causes the high-spin to low-spin ($S=0$ $a_{1g}a_{1g}$ configuration) transition at the doubly occupied sites.
Since the $a_{1g}$ and ${e_g}^\pi$ orbitals are hybridized through inter-pair hopping integrals,
the transition is not as steep as we have seen in the VO$_2$ calculations.
However, a large change in the orbital occupation from $n_{a1g}:n_{eg}\sim 1:1$ to $n_{a1g}:n_{eg}\sim 1:0$
is expected.
 
In conclusion, the MIT in VO$_2$  has been discussed
using the three-band Hubbard model in connection with the lattice distortion on the basis of the many-body theory.
A novel mechanism on the MIT is proposed.
While  the material is considered to be a Mott-Hubbard insulator in the insulating phases, 
the MIT is not induced by an increase in the relative strength of the Coulomb interaction
against the electron hopping, as in the usual Mott transition, 
but by the level splitting between the $\sigma$ and $\pi$ orbitals against the on-site exchange interaction, 
where the charge fluctuation is suppressed  by the change in electron configuration 
in doubly occupied sites from a high-spin ($S=1$)  to a low-spin ($S=0$) state with much larger interaction energy.

This work is partly supported by a Grant-in-Aid for Scientific Research from the Ministry
of Education, Culture, Sports, Science and Technology.

\end{document}